\documentclass[a4paper]{article}

\usepackage{icrc2013}

\title{Lorentz Invariance Violation Limits from the Crab Pulsar using VERITAS}

\shorttitle{ICRC 2013 LIV Limits from Crab PSR}

\authors{
Benjamin Zitzer$^{1}$,
for the VERITAS Collaboration.
}

\afiliations{
$^1$ Argonne National Laboratory, 9700 S. Cass Ave. Lemont, IL, USA \\
}

\email{bzitzer@anl.gov}

\abstract{
Quantum gravity (QG) theories over the past fifty years have sought to understand the relationship 
between the four fundamental interactions. A major insight gained in this area is that all interactions 
could possibly unify at Planck-scale energies $\sim10^{19}$ GeV. A potential consequence of the 
unification of gravity with the other three interactions would be a breaking of Lorentz symmetry at Planck-scale energies.
The interpretation of Time-of-flight (TOF) measurements from gamma-ray telescopes have been able to put constraints on the 
energy scales of the Lorentz-invariance violations (LIV). The Crab pulsar, the only pulsar detected at 
very high energies (VHE, $E>$ 100 GeV) presents a unique opportunity to put new constraints on LIV. 
Presented here are the results of observations of the Crab pulsar with VERITAS and statistical methods 
to determine limits of LIV effects from energy-dependent timing differences.
}

\begin{document}
\maketitle

\section{Introduction}

The Crab Nebula is believed to be the remnant of a supernova observed in 1054 A.D.; nine hundred and fourteen years later, a high-$\dot{E}$ pulsar with a period of $\sim$33 ms was discovered in the system \cite{Crab1968}. 

All 117 known $\gamma$-ray pulsars, including the Crab, show a spectral cutoff above a few GeV \cite{Celik2012}. Traditional pulsar models attribute this cutoff to curvature radiation originating within the magnetosphere. Measuring the spectral break energy and cutoff shape helps to constrain these models. MAGIC's observations of the Crab pulsar revealed significant pulsations at 25 GeV with hints of signal at energies higher than 60 GeV \cite{MagicSci2008}. For the first time, the possibility existed of a non-exponential cutoff in the spectrum of a pulsar. Pulsed emission was later detected by VERITAS above 120 GeV, rejecting the exponential cutoff model at the 5.6$\sigma$ level \cite{OtteSci2011} (see Figure 1).    

\begin{figure}[h]
\includegraphics[width=0.5\textwidth]{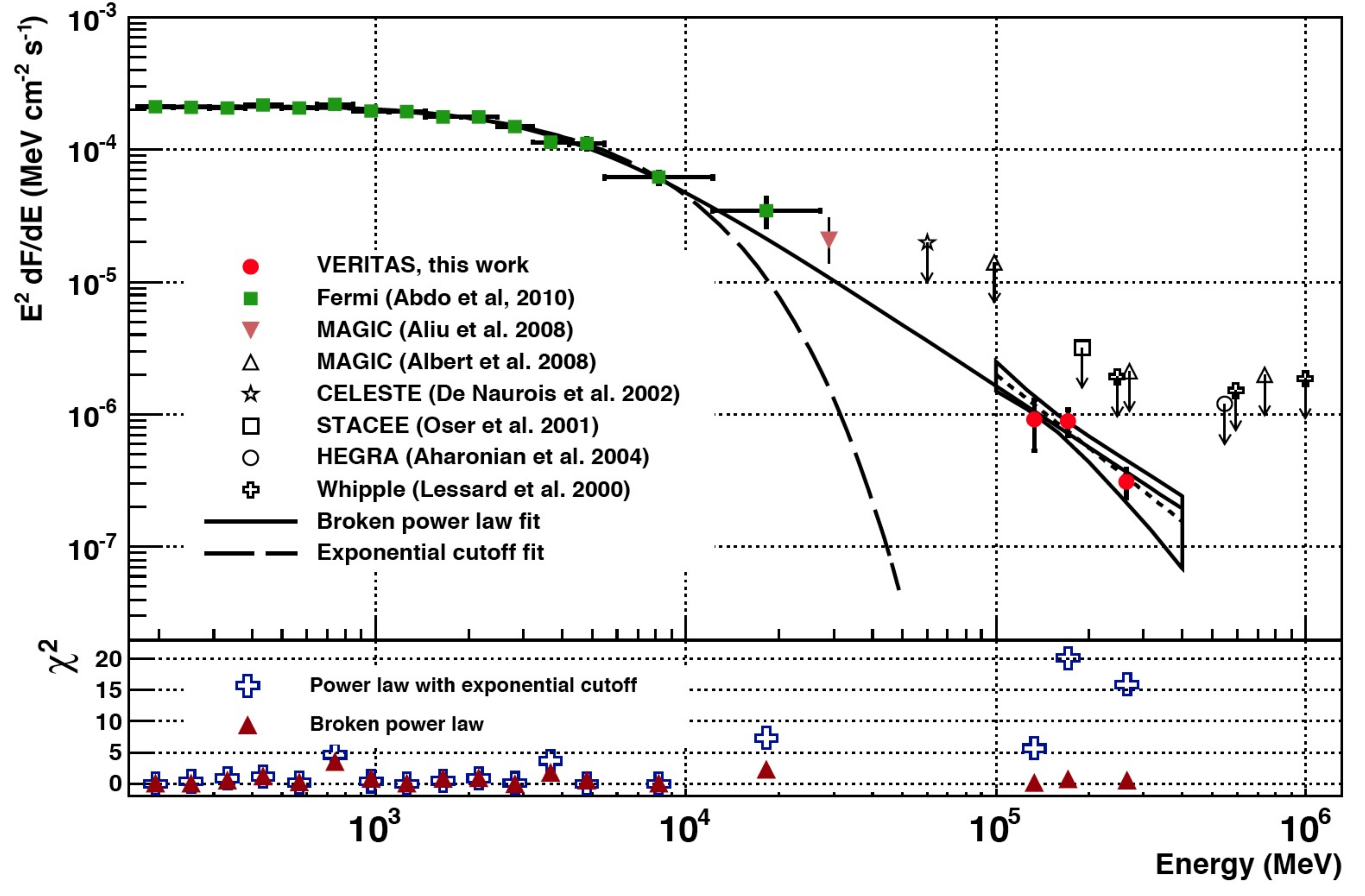}
\caption{Spectral energy distribution of the Crab Pulsar at $\gamma$-ray energies, including VERITAS results. Figure is from \cite{OtteSci2011}. }
\end{figure}

In several quantum gravity models and Standard-Model Extension scenarios, deviations from Lorentz symmetry could emerge from an underlying unified theory. Due to a possible foamy nature of space-time, the speed of light in a vacuum could vary depending on the energy of a particle (for a recent review, see \cite{LIV}). The energy scale for these violations, $E_{QGn}$, could therefore be constrained by time-of-arrival differences between photons of different energy originating from the same source. Any sort of time-of-flight (TOF) testing for these violations would require $\gamma$-ray sources, with fast variability that are seen at astronomical distances, such as AGN, GRBs and the Crab pulsar above 120 GeV. In many cases the speed of light for a photon with energy $E$ can be expanded as:
\begin{equation}
c(E) = c_{0} \left[1 - s_{\pm} \frac{n+1}{2}\left(\frac{E}{E_{QGn}} \right)^{n}\right] ,
\end{equation}
where $E_{QGn}$ is the energy scale where LIV effects are relevant and $c_{0}$ is the speed of light at low energies, 3$\times$10$^{8}$m/s. As an example, for a object a distance $D$ away, if two photons were emitted simultaneously with energies $E_{h}$ and $E_{l}$ with $E_{h} > E_{l}$, the time difference measured by the detector is:
\begin{equation}
\Delta t = s_{\pm}\frac{D}{c_{0}}\frac{E_{h} - E_{l}}{E_{QG1}}
\end{equation}
if the linear term is dominant. $s_{\pm}$ is equal to +1 in the sub-luminal case, and -1 in the super-luminal case. If the quadratic term is dominant, then the timing difference is:
\begin{equation}
\Delta t = s_{\pm}\frac{D}{c_{0}}\frac{E^{2}_{h} - E^{2}_{l}}{E^{2}_{QG2}}.
\end{equation}
Typically either $E_{QG1}$ or $E_{QG2}$ dominates, and TOF measurements have been able to constrain these quantities.  

The Crab Pulsar currently (Spring 2013) presents a unique opportunity for LIV TOF measurements, and there are several reasons why it makes a tempting target for these types of studies:
\begin{enumerate}
\item{The Crab Pulsar's signal to noise ratio increases linearly with time, so limits can be improved by simply observing longer. Tests do not rely on random transient events with low statistics. }
\item{Since the timing of the Crab Pulsar is widely studied throughout the electromagnetic spectrum, energy delays due to propagation effects can be more easily distinguished from intrinsic effects. }
\end{enumerate}

\section{Data Selection and Timing Analysis}

VERITAS (Very Energetic Radiation Imaging Telescope Array System) is an array of four IACTs (imaging atmospheric Cherenkov telescopes) located at the Fred Lawrence Whipple Observatory (FLWO) in southern Arizona (31 40N, 110 57W,  1.3 km a.s.l.) \cite{Galante2012}. VERITAS collected 107 hours of low zenith angle observations on the Crab from the start of four-telescope operations in 2007 through 2011. Data quality selection requires a clear atmospheric conditions, based on infrared sky temperature measurements and nominal hardware operation. Event selection that was applied to the data was optimized \emph{a priori} by assuming a power-law spectrum with an index of -4.0 and a normalization of a few percent of the Crab at 100 GeV \cite{OtteSci2011}. Data reduction followed the standard methods, yielding consistent results with two analysis packages \cite{Analysis}. 

The Jordell Bank timing ephemeris was used to obtain the timing parameters for the pulsar analysis \cite{JordellBank}. Barycentering was done with two custom codes and with tempo2 \cite{Tempo2}. Applying the H-test  \cite{DeJager1994} to this data set yields a H value of 50, corresponding to a 6.0$\sigma$ significance \cite{OtteSci2011}. Defining the significance for pre-chosen ON and OFF regions of the pulse profile according to  Li \& Ma \cite{LiMa1983} gives a 8.8$\sigma$ significance \cite{OtteSci2011}. An unbinned maximum-likelihood fit determined the positions of P1 and P2 to be -0.0023$\pm$0.0020 and 0.0398$\pm$0.002, respectively. The ratio of the number of pulsed events in P2 over the number of pulsed events in P1 is 2.4$\pm$0.6 \cite{OtteSci2011}. The pulse profiles measured by VERITAS and the \emph{Fermi}-LAT is shown in Figure 2.  

\begin{figure}[h]
\includegraphics[width=0.5\textwidth]{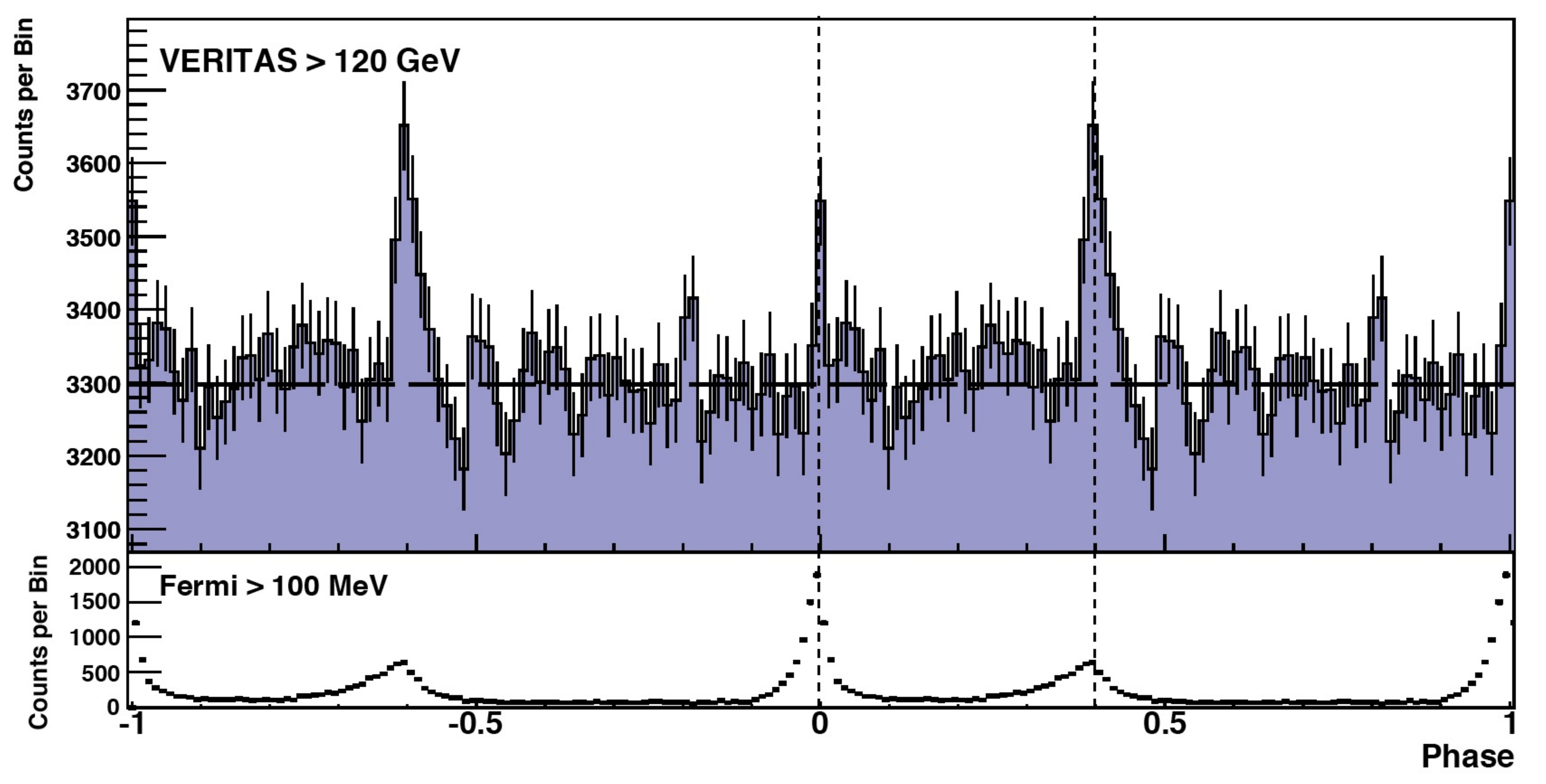}
\caption{Pulse profile of the Crab pulsar at $\gamma$-ray energies with VERITAS. All quality data between 2007 and 2011 is included, the exact data set used for \cite{OtteSci2011}. The \emph{Fermi}-LAT pulse profile is also shown below the VERITAS pulse profile). }
\end{figure}

\section{Methodology and Results}

\subsection{Peak Timing Comparison}

The pulse profile of the VERITAS data above 120 GeV is compared to the pulse profile of the \emph{Fermi}-LAT data above 100 MeV. If the same timing solutions are used for both data sets, then the peak positions agree within statistical uncertainty. This indicates no measurable violations of Lorentz invariance, so a lower limit on $E_{QGn}$ is  therefore calculated, using Equations 2 and 3. The 95\% confidence upper limit on the timing of the peaks is calculated to be less than 100 $\mu$s. The limits of the linear LIV term is therefore:
\begin{equation}
E_{QG1}> \frac{d \Delta E}{c_{0}  \Delta t_{95\%}} = \frac{2\mathrm{kpc} * 120 \mathrm{GeV}}{3\times10^{8} \mathrm{m/s} * 100 \mathrm{\mu s} } \sim 3\times10^{17} \mathrm{GeV} 
\end{equation} 

\subsection{Dispersion Cancellation}

The method described in the previous section relies on binning the data in both energy and in pulsar phase. Techniques involving binning always involve a loss of information. Additionally, binning in energy is not ideal because of the variations of the pulse period within the energy bins due to pulsar spin-down. The ideal methodology for LIV, if possible, should be unbinned in both energy and time (or phase in this case). The large $\gamma$-ray background due to the Crab Nebula provides additional problems. This section discusses a variation of the Dispersion Cancellation (DisCan) method \cite{Scargle2008,FermiGRB2009}, that is well-suited to use for pulsars.

The method here utilizes the $Z^{2}_{m}$ test \cite{Buccheri1982} as a test statistic. The $Z^{2}_{m}$ test is derived from a Fourier-series estimator which tests for variations from a uniform (unpulsed) light curve for a chosen Fourier harmonic $m$ \cite{DeJager1986}. $Z^{2}_{m}$ is proportional to the Fourier power of the pulsar. LIV effects would introduce a dispersion of the pulsar signal. The maximal value of $Z^{2}_{m}$, therefore, corresponds to the Fourier power of the undispersed signal. $Z^{2}_{m}$ takes the form:
\begin{equation}
Z^{2}_{m}=\frac{2}{N}\sum^{m}_{j=1}[{(\sum^{N}_{i=1}{\sin(2\pi\phi_{i}j)})^{2}} + {(\sum^{N}_{i=1}{\cos(2\pi\phi_{i}j)})^{2}}],
\end{equation}
where $N$ is the total number of events and $\phi_{i}$ is the phase of the $i^{th}$ event (mod 1). 

The procedure used is as follows:

\begin{enumerate}
\item{Adopt a model for a correction to the arrival time of each event, as a function of the event energy. For example, if the LIV effect has the form of $E^{n}$, then the correction for an event of arrival time of $t_{i}$ and energy $E_{i}$ is:
\begin{equation}
t'_{i}=t_{i}-\theta E^{n}_{i}
\end{equation}
}
\item{Refold the pulsar phases according to the formula above for a choice of $\theta$. 
\begin{equation}
\phi'_{i}=(t'_{i}-t_{0,i})\nu_{i} + \frac{1}{2}(t'_{i}-t_{0,i})^{2}\dot{\nu}_{i} (\bmod 1),
\end{equation}
where $t_{0,i}$ is the pulsar epoch, and $\nu$, $\dot{\nu}$ is the pulsar frequency and 1st derivative of the pulsar frequency, respectively. $\theta$ could hypothetically be any real number, positive or negative, but with some common sense it can be narrowed down. LIV effects are small at GeV/TeV scales, not significant enough to drastically change the intrinsic shape of the pulse profile. The pulsed spectrum of the Crab pulsar extends to $\sim$400 GeV. It is therefore unlikely that a photon at 400 GeV will move in phase more than 5\% of the pulse period of Crab in either direction due to LIV effects. This limits the range in $\theta$ to $|$ 400 GeV/(0.05* 33 ms) $|$ = $|\theta|$ $ < 4.1 \mu$s/GeV.
}
\item{Calculate $Z^{2}_{m}$ as above in equation 5.}
\item{Repeat steps 2 and 3 for several test values of $\theta$, finding a value of $Z^{2}_{m}$ that is maximized ($\theta_{max}$).}
\item{Repeat steps 2 through 4 on several bootstrapped or Monte Carlo (MC) data sets to determine a probability distribution function (PDF) that will determine test significance and limits.
}

\end{enumerate}

In addition to being unbinned in both phase and energy, this approach uses \emph{all} photons (P1, P2 and background events), removing all potential trial factors except from for choice of $m$, which is determined to be 20 from a MC optimization. Results of MC tests of this approach are shown in Figure 3. It should be noted that 20 was the optimal value $\it{only}$ investigating the pulse profile taken with VERITAS from \cite{OtteSci2011}. Other pulse profiles from other pulsars or even an updated pulse profile for the Crab with data taken after 2011 could change the optimal value. 

\begin{figure}[t]
\includegraphics[width=0.5\textwidth]{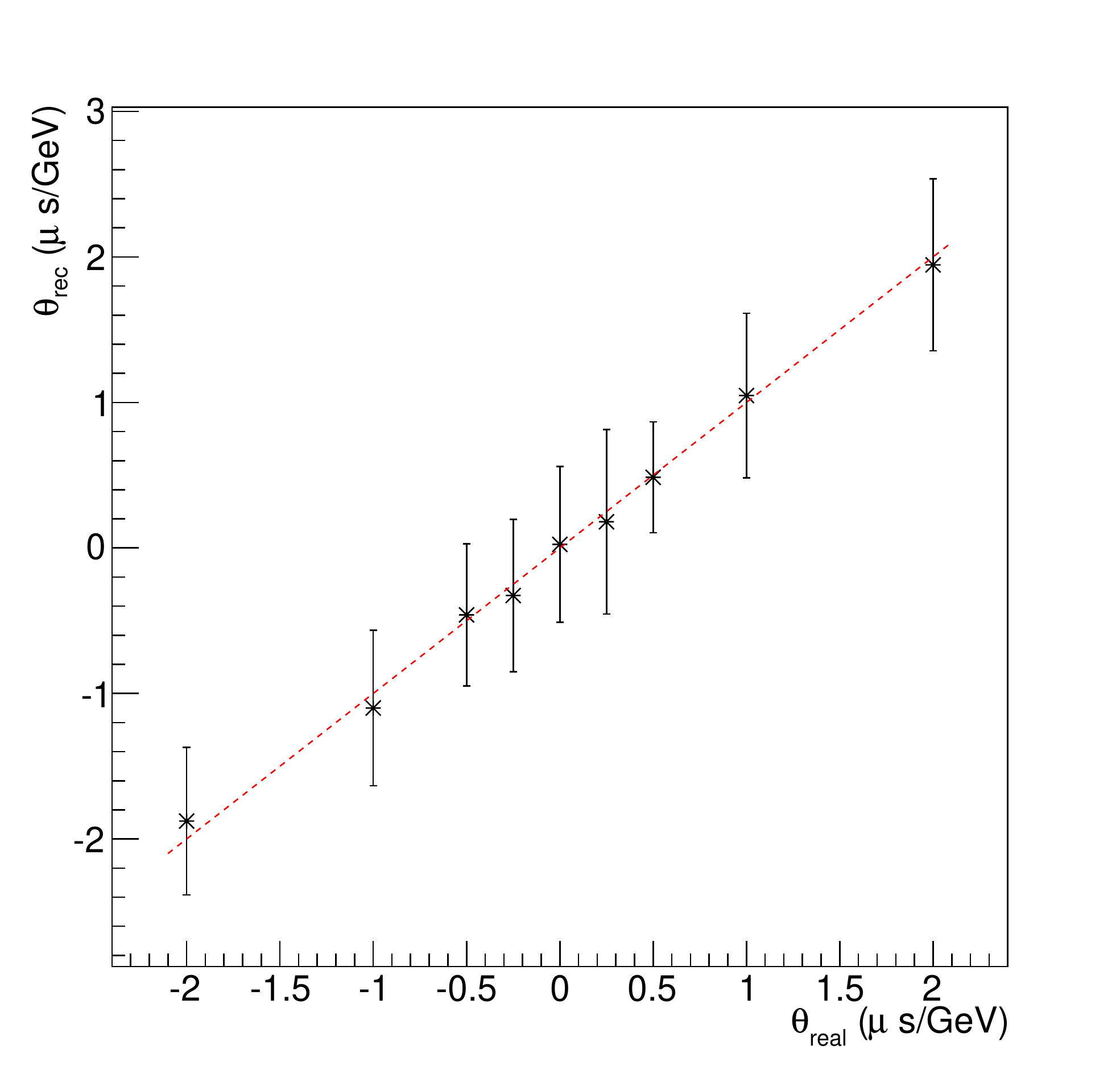}
\caption{Performance of the $Z^{2}_{20}$ DisCan results on a toy MC. For each value of $\theta_{real}$, one hundred realizations of the energy distribution is produced and event times are shifted by equation 6 with $n$=1. The DisCan algorithm is applied to each MC to recover the value of $\theta_{real}$. The error bars of $\theta_{rec}$ is the RMS of the one hundred realizations. The red dashed line represents x=y.  }
\end{figure}

The $Z^{2}_{20}$ DisCan test, when applied to the data set from the 2007 to 2011 seasons, has a maximal value at $\theta$ = -0.49 $\mu$s/GeV. The of plot $Z^{2}_{20}$ against trial values of $\theta$ is shown in Figure 4. To determine the statistical significance and limits of this test, a PDF is produced by one thousand MC realizations of the energy distribution. The $Z^{2}_{20}$ DisCan test is applied to each one. The distribution of the $\theta_{max}$ values is the PDF. Figure 5 shows the PDF produced. The maximum $\theta$ found in the data, -0.49$\mu$s/GeV, has a significance of 1.4$\sigma$ away from the null result of $\theta$=0.

\begin{figure}[h]
\includegraphics[width=0.5\textwidth]{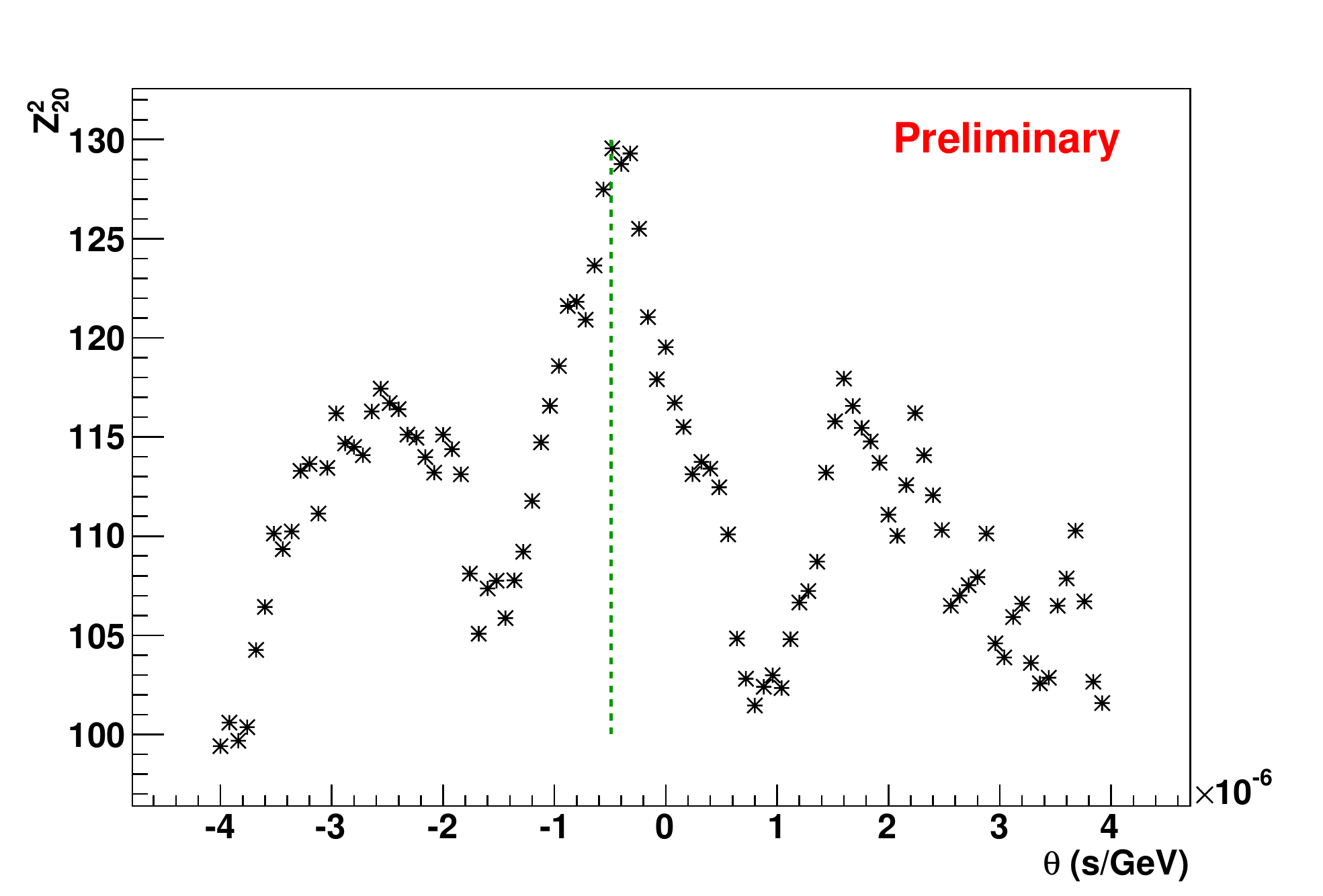}
\caption{Preliminary performance of the $Z^{2}_{20}$ DisCan method on the VERITAS Crab Pulsar data between 2007 and 2011. Trial values of $\theta$ are plotted against $Z^{2}_{20}$ The green dashed line is the maximum Z$^{2}$ value at $\theta_{max}$ = -0.49$\mu$s/GeV. }
\end{figure}

The LIV energy scale is related to $\theta$ by:
\begin{equation}
E_{QGn}=(s_{\pm}D/c_{0}\theta)^{1/n}.
\end{equation}
To place lower and upper bounds on $\theta$, Bayes theorem was used to determine the cumulative posterior PDF with the likelihood PDF derived from MC simulations shown in figure 5. It is assumed that the shape of the likelihood PDFs is independent on the value of $\theta$ used in the simulations \cite{Vasileiou2013}. With this method, 95\% confidence limits for $\theta$ of -1.2 $\mu$s/GeV and 1.1 $\mu$s/GeV were derived for the lower and upper limits, respectively. The Crab pulsar is located 2 kpc away, giving a sub-luminal limit of the linear energy scale of $E_{QG1} >$1.9$\times$10$^{17}$ GeV and a super-luminal limit of $E_{QG1} >$1.7$\times$10$^{17}$ GeV.  

\section{Conclusions}

This work presented two very different methods for measuring LIV from the Crab Pulsar which both yielded similar limits. The limits obtained from the peak timing differences here are comparable to limits found from MAGIC with Mrk 501 data \cite{MagicLIVLim}, an order of magnitude below limits found with AGN from HESS \cite{HESSLIVLim} and less than two orders of magnitude below the Planck mass scale. The dispersion cancellation method showed a possible hint of towards the super-luminal case. It should be noted that the bounds determined by the dispersion cancellation are likely to have large errors associated with them, due to a small number of statistics in the tails in the probability distribution in figure 5. The method discussed here can be improved in the future by using greater than one thousand trials for the probability distribution.   

\begin{figure}[h]
  \centering
  \includegraphics[width=0.5\textwidth]{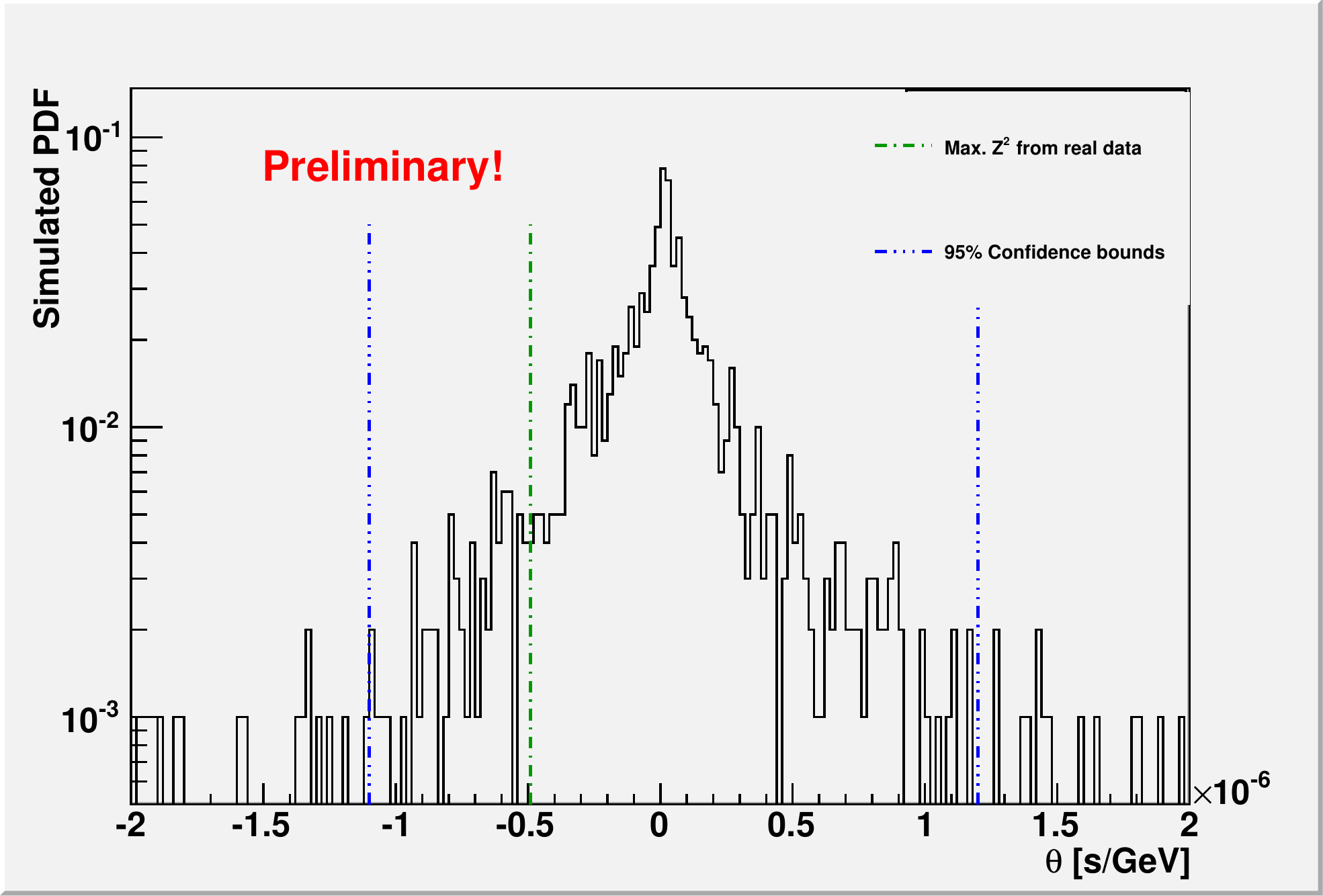}
  
  \caption{Probability distribution function generated with one thousand realizations of the energy distribution used to extract limits. The green dashed line is at the value of $\theta_{max}$.  The blue dashed lines represent where bounds were placed.}
  \label{wide_fig}
 \end{figure}

 While the Crab Pulsar does not $\it{currently}$ have the best sensitivity to LIV measurements, there is still merit to the result. Some postulate that LIV effects would not be isotropic \cite{Kostelecky}, so multiple results from different targets could constrain the anisotropy of the effect. Finally, it is important to get measurements at multiple distances to completely eliminate any sort of intrinsic effect, since they would not be dependent on redshift.

It is possible that the Crab is not unique as a VHE emitting pulsar, and the methods mentioned here could certaintly be used for any pulsar. Hypothetically, if a millisecond pulsar was discovered at VHE energies, it would have $\sim$10 times the frequency, and therefore $\sim$10 times the LIV sensitivity to the linear term.   

As mentioned earlier with more observing time this limit could be greatly improved. A small improvement could be gained by adding Fermi-LAT data to the DisCan method, although the method is more sensitive at higher energies. Additional data will improve the signal-to-noise of the pulse profile which will improve the limits, as well as extend the spectrum of the pulsar to higher energies if the power-law trend continues. Additionally, re-analysis of the data using the DisCan method with different cuts that provide better energy resolution could improve the limit by as much as a factor of $\sim$2. 

\vspace*{0.5cm}
\footnotesize{{\bf Acknowledgment:}{This research is supported by grants from the U.S. Department of Energy Office of Science, the U.S. National Science Foundation and the Smithsonian Institution, by NSERC in Canada, by Science Foundation Ireland (SFI 10/RFP/AST2748) and by STFC in the U.K. We acknowledge the excellent work of the technical support staff at the Fred Lawrence Whipple Observatory and at the collaborating institutions in the construction and operation of the instrument. The author would like to thank P. Kaaret, S. Griffiths, N. Otte and A. McCann for their productive conversations and assistance. }}


\begin{thebibliography}{}

\bibitem[1]{Crab1968} D. Staelin et al., \emph{Science} 162, 3861 (1968)
\bibitem[2]{Celik2012} O. Celik, \emph{Gamma 2012 Proceedings} (2012)  
\bibitem[3]{MagicSci2008} E. Aliu et al., \emph{Science} 322, 1221 (2008)
\bibitem[4]{OtteSci2011}  E. Aliu, et al., \emph{Science} 334, 69 (2011) 
\bibitem[5]{LIV}N. Otte, et al., arXiv:1305.0264 (2013)
\bibitem[6]{Galante2012} J. Holder et al., \emph{Astroparticle Physics} 25, 391 (2006)
\bibitem[7]{Analysis} V.A. Acciari et al. \emph{ApJ} 679, 1427 (2008)
\bibitem[8]{JordellBank} A. G. Lyne et al., \emph{MNRAS} 265, 1003 (1993)
\bibitem[9]{Tempo2} G. B. Hobbs, R. T. Edwards and R. N. Manchester, \emph{MNRAS} 369, 655 (2006)
\bibitem[10]{DeJager1994} O. C. de Jager, \emph{ApJ} 436, 239 (1994)
\bibitem[11]{LiMa1983} Li, T.-P. and Ma, Y.-Q., \emph{ApJ} 272, 317 (1983)
\bibitem[12]{OtteICRC2011} N. Otte, \emph{ICRC 2011 Proceedings} 7, 255 (2011) 
\bibitem[13]{Scargle2008} J. D. Scargle et al., \emph{ApJ} 673, 972 (2008)
\bibitem[14]{FermiGRB2009} Abdo, A.~A., Ackermann, M., Ajello, M., et al., \emph{nat}, 462, 331 
\bibitem[15]{Buccheri1982}Buccheri, R., Bennett, K., Bignami, G.~F., et al., \emph{aap}, 128, 245 (1983) 
\bibitem[16]{DeJager1986}de Jager, O.C., Raubenheimer, B.~C., \& Swanepoel, J.~W.~H. \emph{aap}, 170, 187 (1986)
\bibitem[17]{Vasileiou2013}Vasileiou, V., Jacholkowska, A., Piron, F., et al., \emph{prd}, 87, 122001 (2013)
\bibitem[18]{MagicLIVLim} Albert, J., et al., \emph{Physics Letters B}, 668 (2008) 
253 
\bibitem[19]{HESSLIVLim}J. Bolmont, A. Jacholkowska, R. Buehler, et al. \emph{38th COSPAR Scientific 
Assembly} 38, 2271 (2010) 
\bibitem[20]{Kostelecky} V.~A. Kosteleck{\'y} \& M. Mewes, \emph{apjl}, 689 (2008)

\end{thebibliography}
\end{document}